\documentclass[aps,pre,showpacs,twocolumn]{revtex4}
\usepackage{graphicx}

\begin{document}
\title{Two-dimensional projections of an hypercube}

\author{Guillermo Abramson}
\email{abramson@cab.cnea.gov.ar} \affiliation{Consejo Nacional de
Investigaciones Cient{\'\i}ficas y T{\'e}cnicas \\ Centro
At{\'o}mico Bariloche and Instituto Balseiro, 8400 S. C. de
Bariloche, Argentina}

\author{Dami{\'a}n H. Zanette}
\email{zanette@cab.cnea.gov.ar}
\affiliation{Consejo Nacional de
Investigaciones Cient{\'\i}ficas y T{\'e}cnicas \\ Centro
At{\'o}mico Bariloche and Instituto Balseiro, 8400 S. C. de
Bariloche, Argentina}

\begin{abstract}
We present a method to project a hypercube of arbitrary dimension
on the plane, in such a way as to preserve, as well as possible,
the distribution of distances between vertices. The method relies
on a Montecarlo optimization procedure that minimizes the squared
difference between distances in the plane and in the hypercube,
appropriately weighted. The plane projections provide a
convenient way of visualization for dynamical processes taking
place on the hypercube.
\end{abstract}


\maketitle

Dynamical models where the state of a system is represented by an
ordered array of binary variables are ubiquitous in statistical
physics, especially in its interdisciplinary applications.
Perhaps the widest class of models that admit such Boolean-like
representation is constituted by binary cellular automata
\cite{cel,kauf0}. Specific applications include biological
evolution at the levels of molecules \cite{volkenstein}, cells
\cite{kauf}, individuals \cite{Penna} and species \cite{KJ,SM},
as well as social and socioeconomical behavior \cite{Palmer,Weis}.
Moreover, genetic algorithms are typically applied to systems
whose configuration is described by means of binary sequences
\cite{GA}. These models may involve large populations of
interacting agents, each of them described as a time-dependent
array of bits, which requires assigning an evolving density to
each possible binary sequence.

While the configuration space of a binary sequence of length $L$
is naturally represented as the set of $2^L$ vertices of an
$L$-dimensional hypercube, its visualization can be
dissapointingly difficult, even for $L$ not very large. On the
other hand, besides a quantitative characterization of the system
dynamics through its collective properties, it is sometimes
desirable to rely on a geometrical depiction where the dynamics
can be followed, for instance, on the computer screen. The
purpose of this paper is to present a method to project the
vertices of a hypercube of arbitrary dimension onto a set of
points in the plane, with the condition of preserving, as much as
possible, the structure of the distance distribution on the
hypercube. The motivation of this condition is that many dynamical
processes depend on the Hamming distance---i.e., the number of
different bits---between binary sequences, and we require this
feature to be well represented by the Euclidean distance between
the corresponding points in the plane projection.

Let $h_{ij}$ be the Hamming distance between vertices $i$ and $j$
in the hypercube, and $d_{ij}$ the Euclidean distance between
points $i$ and $j$ in their plane projection. We define the
function
\begin{equation}
E=\sum_{i,j}(d_{ij}-h_{ij})^2,
\label{energy1}
\end{equation}
that characterizes how different are the distances between pairs
of vertices and their projections. Our goal is to find a plane
distribution that minimizes $E$, thus optimizing the plane
representation of the hypercube with respect to the distance
between pairs. We have implemented a Montecarlo method to
approach stochastically the optimal solution---the configuration
of minimum ``energy'' $E$. Starting from a random initial
configuration on the plane, each point performs a walk with fixed
step length $r$ and  directions chosen at random with uniform
probability in $[0,2\pi)$. Each step of this walk produces a
change in the configuration and, hence, in the distances $d_{ij}$,
which implies a variation $\Delta E$ in the energy. The new
configuration is accepted with probability
\begin{equation}
p= \left\{
\begin{array}{ll}
\exp (-\Delta E/T) & \mbox{if $\Delta E>0$} \\
1 & \mbox{otherwise},
\end{array}
\right.
\end{equation}
and rejected with probability $1-p$. The ``temperature'' $T$
parametrizes this probability and allows the usual implementation
of a simulated annealing, where the procedure starts with a high
temperature that enables the system to explore a wide range in
configuration space. Progressively, the temperature is reduced
and the system freezes in one of the many local minima of the
energy, typically not far away from the global minimum if the
annealing is made slowly enough.

\begin{figure}[tbp]
\centering
\resizebox{\columnwidth}{!}{\includegraphics{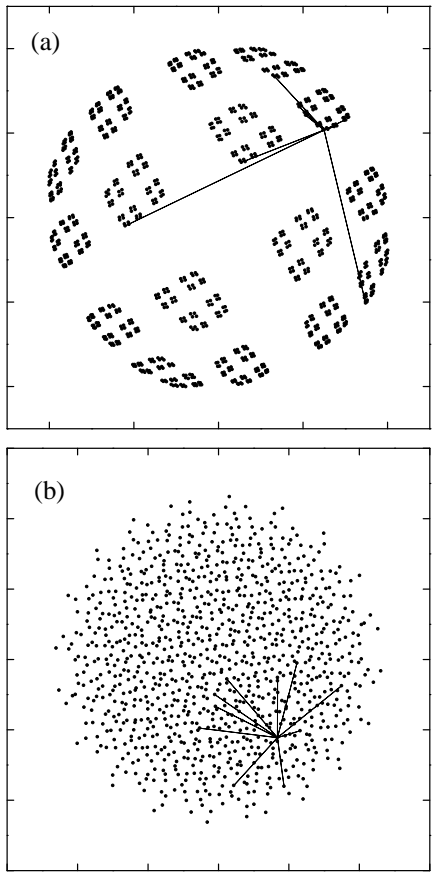}}
\caption{Two plane projections of a $10$-dimensional hypercube.
(a) Using the energy defined in Eq.~(\ref{energy1}). (b) Using the
energy defined in Eq.~(\ref{energy2}). Lines join the projections
of a randomly chosen vertex of the hypercube and its nearest
neighbors.} \label{projections1024}
\end{figure}

We have carried out the described procedure both interactively,
reducing by hand the temperature while monitoring the
configuration of the system on the computer screen, and
automatically, by implementing a programmed reduction of the
temperature. Our experiments show that essentially the same state
is achieved in almost all the realizations. This implies that the
energy landscape, while rugged, does not posses deep local minima
that could capture the configuration far from the optimal one.
The typical final configuration for $L=10$ ($N=1024$ points) is
shown in Fig.~\ref{projections1024}(a). The self-similarity of
its structure is remarkable, since no such property is present in
the hypercube. Despite the appeal that this self-similar
projection may have, it turns out that such projection is not well
suited for our purpose. Vertices that are relatively near in the
hypercube result rather far away in the projection. As an
illustration, the first neighbors $j$ ($h_{ij}=1$) of a given
vertex $i$ are shown in the figure. It is apparent that the
Euclidean distance of some of them from the reference vertex is
comparable with the size of the system. Moreover, many other
vertices which should be farther away from vertex $i$, are in fact
much closer.

One way to solve this difficulty is to modify the definition of
$E$, such that near neighbors have more weight than distant
neighbors. In fact, the energy (\ref{energy1}) overemphasizes the
effect of large distances. We have implemented the following
simple alternative:
\begin{equation}
E=\sum_{i\neq j}\left(\frac{d_{ij}-h_{ij}}{h_{ij}}\right)^2.
\label{energy2}
\end{equation}
The final configuration, which we will term ``homogeneous,'' is
shown in Fig.~\ref{projections1024}(b). Neighbor vertices of a
vertex $i$ now result mapped onto points that surround the point
$i$, which makes this projection much more satisfying. Certainly,
however, some vertices result mapped near the border of the
circle, and the arrangement of their neighbors is slightly
different than that of vertices mapped in the middle of the set.
We analyze below how this affects the distribution of distances.

\begin{figure}[tbp]
\centering
\resizebox{\columnwidth}{!}{\includegraphics{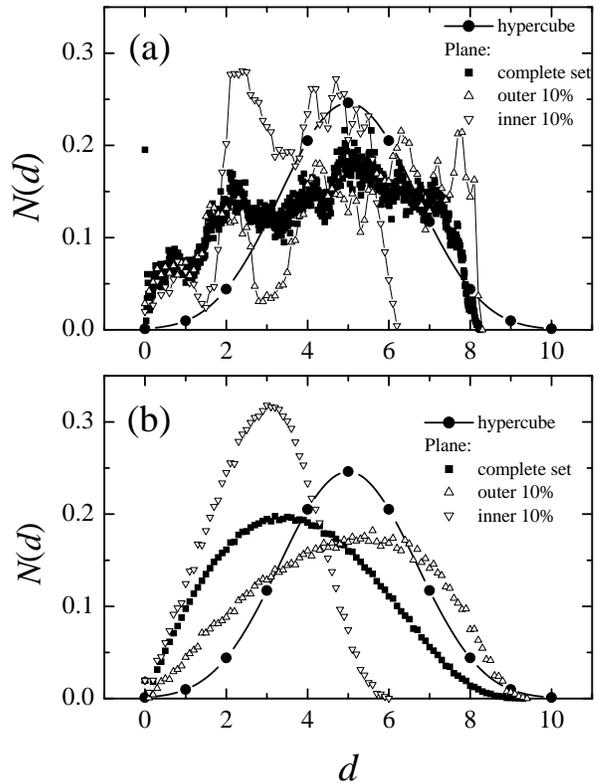}}
\caption{Distribution of distances in the hypercube, the plane
projections, and selected subsets of these. (a) Self-similar. (b)
Homogeneous.} \label{distances}
\end{figure}

A good characterization of the projections, and a quantitative
way for comparing them, is the distribution of distances in each
set. In the hypercube the distribution of distances to a vertex
is the same for every vertex, and in fact is analytically found
to be a binomial distribution. In the two-dimensional projections
there is a different distribution for each point of the set.
In Fig.~\ref{distances}(a) and (b) we show normalized
distributions of distances for the self-similar and the
homogeneous projections, respectively. In both figures, the black
circles show the distribution of the distance to (any) vertex in
the hypercube. Even though the distances form a discrete set, we
show lines connecting the points to ease the reading of the
graph. The other three curves shown in each plot correspond to
the plane projections. The black squares correspond to an average
on all the points in the sets. Triangles show averages performed
on either the 10\% of the points that form the external corona of
the projection, or the 10\% of its more central points. For the
points of these subsets, still, all the distances to other points
of the whole set are taken into account in the distributions.

The most immediate observation regarding Fig.~\ref{distances} is
the difference between the distributions in the two projections.
The self-similar projection displays rugged distributions that
reflect the hierarchical geometrical arrangement of the points.
In the homogeneous projection, instead, the distributions are
smooth, as in the hypercube. To this extent, the homogeneous
projection can be said to represent more accurately the
distribution of distances present in the hypercube. The
distribution averaged over the whole set appears, however, skewed
towards smaller distances, with a maximum around $d=3$, instead
of the most represented distance $h=5$ of the hypercube.
Interestingly, Fig.~\ref{distances}(b) shows that the
\emph{outer} 10\% points considerably correct this skew. In other
words, a point near the border of the circular array of the
projection has a distribution of distances to the other points in
the set which is rather similar to the distribution of a vertex
of the hypercube.

\begin{figure}[tbp]
\centering
\resizebox{\columnwidth}{!}{\includegraphics{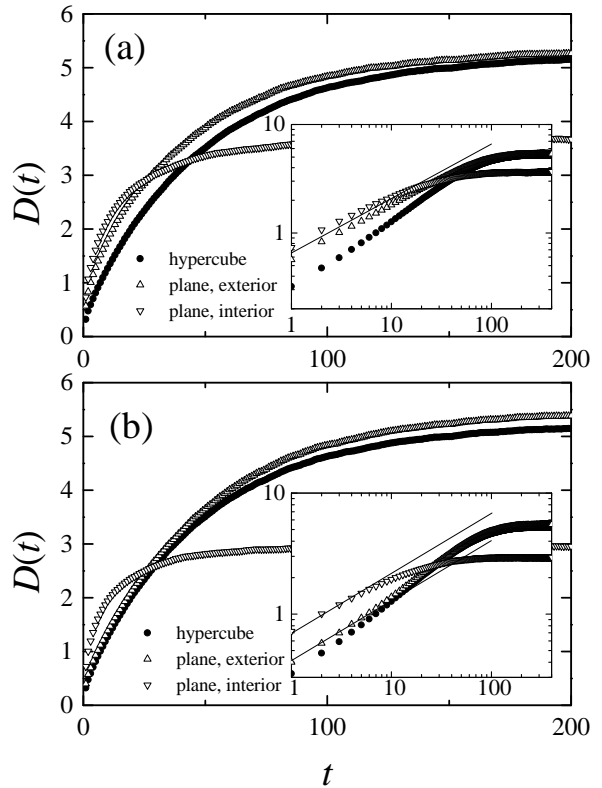}}
\caption{Average displacement as a function of time for diffusion
in the hypercube and its plane projections. In these last,
starting from a point in the border and a point in the center. (a)
Self-similar. (b) Homogeneous. The staright lines in the insets
have slope $1/2$.} \label{diffusion}
\end{figure}

An appraisal of the plane projections of the hypercube in a
dynamical context results from the consideration of a diffusion
process. Let us suppose that, at each time step, a random walker
jumps from a vertex of the hypercube to one its neighbors with
equal probability. The average distance $D$ from the initial site,
as a function of time, is shown in Fig.~\ref{diffusion}(a) and
(b) as black circles. The inset in both figures displays the same
curve in double logarithmic scales, showing an initial behavior
of the form $D(t) \sim t^{1/2}$, like in a regular random walk in
Euclidean space, followed by a saturation as the hypecube space
is fully explored. The average distance as measured on the plane
is shown in Figs.~\ref{diffusion}(a) and (b) for the self-similar
and the homogeneous projections respectively. As expected from the
distance distribution discussed above, the results for the plane
projections depend on whether the initial point of the walker is
at the border or at the center of the set. These two cases are
shown in Figs.~\ref{diffusion}(a) and (b) as triangles pointing
upward and downward, respectively. From this dynamical point of
view, interior points behave equally bad in both projections. The
most faithful representation of the process in a plane projection
is the one given by one of the border points of the homogeneous
set (Fig.~\ref{diffusion}(b), up triangles). Diffusion starting
at these points behaves similarly as from points of the
hypercube, both in the short and in the long time regimes, as seen
in the linear and the logarithmic plots.

Our main goal of obtaining a sensible plane projection of the
hypercube with the purpose of visualizing a dynamical process has
been achieved, to an acceptable extent, by the homogeneous
projection. Suppose that a dynamical phenomenon is taking place
in a neighborhood of vertex $P$ of a hypercubical phase space. We
need to build a homogeneous projection that maps vertex $P$ to a
point at the border of the plane set. This is easily done by
generating a projection at random and identifying one of the
points at the border first. Suppose that one such point is $Q$.
Then, each vertex $I$ of the hypercube is mapped to a point in the
plane projection as
\begin{equation}
I\rightarrow (I \oplus P) \oplus Q,
\end{equation}
where $\oplus$ stands for the bitwise exclusive-or (XOR) operator.
The projection obtained in this way provides a nice plane
visualization substrate for the process.

\end{document}